\documentclass[prd,twocolumn,twoside,preprintnumbers,superscriptaddress,nofootinbib]{revtex4}

\usepackage{amsmath}
\usepackage{graphicx}
\usepackage{dcolumn}
\usepackage{bm}
\usepackage[hyperfootnotes=false]{hyperref}
\usepackage{xspace}

\newcommand{\Fpi}{F_\pi}
\newcommand{\mpi}{M_{\pi}}
\newcommand{\mpii}{M_{\pi^0}}
\newcommand{\ga}{g_{\rm A}}
\newcommand{\Order}{\mathcal{O}}
\newcommand{\muu}{m_u}
\newcommand{\md}{m_d}
\newcommand{\mpp}{m_p}
\newcommand{\mn}{m_n}
\newcommand{\MeV}{\,\text{MeV}}

\begin{document}

\title{Accurate evaluation of hadronic uncertainties in spin-independent \\ WIMP--nucleon scattering: Disentangling two- and three-flavor effects}

\author{Andreas Crivellin}
\author{Martin Hoferichter}
\author{Massimiliano Procura}
\affiliation{Albert Einstein Center for Fundamental Physics,
Institute for Theoretical Physics,\\ University of Bern, Sidlerstrasse 5, CH--3012 Bern, Switzerland}

\begin{abstract}
We show how to avoid unnecessary and uncontrolled assumptions usually made in the literature about soft $SU(3)$ flavor symmetry breaking in determining the two-flavor nucleon matrix elements relevant for direct detection of WIMPs. Based on $SU(2)$ Chiral Perturbation Theory, we provide expressions for the proton and neutron scalar couplings $f_u^{p,n}$ and $f_d^{p,n}$ with the pion--nucleon $\sigma$-term as the only free parameter, which should be used in the analysis of direct detection experiments. This approach for the first time allows for an accurate assessment of hadronic uncertainties in spin-independent WIMP--nucleon scattering and for a reliable calculation of isospin-violating effects. We find that the traditional determinations of $f_u^p-f_u^n$ and $f_d^p-f_d^n$ are off by a factor of $2$.
\end{abstract}
%

\maketitle

\section{Introduction}
\label{intro}

Establishing the nature of dark matter (DM) is one of the fundamental open problems in particle physics and cosmology. A weakly interacting massive particle (WIMP) is an excellent candidate since, for masses in the GeV to TeV range, it naturally provides a relic abundance consistent with that required of DM. 
Direct detection experiments aim at measuring recoil energy depositions in WIMP scattering on a nuclear target with highly sensitive detectors. Claims of a signal by DAMA~\cite{Bernabei:2008yi}, and excess events by CoGeNT~\cite{Aalseth:2011wp}, CRESST~\cite{Angloher:2011uu}, and CDMS II~\cite{Agnese:2013rvf} have been contested by null observations by XENON~\cite{Aprile:2011hi,Angle:2011th} and LUX~\cite{Akerib:2013tjd}. 
In order to fully exploit constraints from present and future measurements (see~\cite{Snowmass} and references therein) and to firmly establish the existence of possible tensions between them, it is crucial to accurately evaluate hadronic uncertainties. Effective field theories (EFTs) provide powerful tools to reach this goal. First of all, effective operators describing the interaction between DM and Standard Model (SM) particles can be organized according to their mass dimension. In the fermionic case, these have the generic schematic structure
\begin{align}
O = \bar{\chi} \Gamma_\chi \chi\; \bar{\psi} \Gamma_\psi \psi
\end{align}
in terms of bilinears built with the DM $\chi$-field and SM $\psi$-fields and $\Gamma \in \{ 1, \gamma^5, \gamma^\mu, \gamma^\mu \gamma^5, \sigma^{\mu \nu} \}$, and  analogously for bosonic operators. Here we focus on spin-independent (SI) interactions since coherence effects lead to an enhancement which is proportional (in the isospin symmetric case) to the square of the number of nucleons in the target nucleus, which is typically heavy. Spin-dependent or momentum-suppressed interactions are much less stringently constrained by direct detection experiments. In formulating theory predictions for SI cross sections, the nucleon matrix elements whose uncertainties play a fundamental role are those involving the
 quark scalar operator $O_{qq}^{SS}$ and the gluon operator $O_{gg}^S$ from the dimension-7 effective Lagrangian 
\begin{equation}
\label{Lagr}
{\cal L}_{\text{eff}}^{(7)} = C_{q q}^{SS} \; \frac{{{m_q}}}{{{\Lambda ^3}}}\;\bar{\chi} \chi \,\bar{q} q + C_{g g}^S\; \frac{\alpha_s}{{{\Lambda ^3}}}\,\bar{\chi} \chi \,G_{\mu \nu} G^{\mu \nu},
\end{equation}
where $q$ denotes quarks fields, $\alpha_s$ the strong coupling, and $G_{\mu\nu}$ the QCD field strength tensor. At the hadronic scale of direct detection experiments, only the light quarks ($u$, $d$, and $s$) and the gluons are active degrees of freedom.
The dimensionless Wilson coefficients $C_j^i$ encode unresolved dynamics at energy scales higher than the cutoff $\Lambda$, which is of the order of the mass of the lightest high-energy particles that get integrated out.

In this paper we stress a point that has been overlooked in the literature and investigate its important implications. Information on nucleon matrix elements involving just $u$- and $d$-quarks have so far been extracted from an empirical formula based on soft flavor $SU(3)$ symmetry breaking~\cite{Cheng:1988im}. This prevents the possibility to assign any reliable theory uncertainty to these predictions. Here we show how to properly relate two-flavor dependent quantities to phenomenology in a rigorous, model-independent way based on Chiral Perturbation Theory (ChPT), the effective field theory of QCD at low energies. In particular, we disentangle two-flavor observables from matrix elements involving the strange quark, which can be more reliably determined from lattice QCD computations.
We clarify the role of the input parameters in the SI WIMP--nucleon cross section in such a way that hadronic uncertainties can now be accurately assessed. While the impact of the pion--nucleon $\sigma$-term $\sigma_{\pi N}$ has been emphasized before~\cite{Bottino:1999ei,Bottino:2001dj,Ellis,HillSolon},
here we work out its effects devoid of unnecessary $SU(3)$ assumptions.
Better convergence is a distinctive feature of the two-flavor chiral expansion in $M_\pi/\Lambda_\chi$ as compared to its three-flavor analog, which involves $M_K/\Lambda_\chi$ corrections, with $\Lambda_{\chi} \simeq 1$ GeV the typical scale of chiral symmetry breaking. 
Moreover, starting from ChPT in its $SU(2)$ formulation allows for the well-controlled calculation 
of isospin-breaking effects, whose incorporation is crucial in the context of isospin-violating DM~\cite{Kurylov:2003ra,Giuliani:2005my,Chang:2010yk,Feng:2011vu,Cirigliano:2012pq,Cirigliano13}. Since the dependence on $\sigma_{\pi N}$ drops out in the difference between proton and neutron couplings, it is here that the shortcomings of the previous prescription become most apparent.  

In the next sections we provide all the formulae that should be used in phenomenological analyses, provide updated expressions for the scalar couplings to $u$- and $d$-quarks, and illustrate the role of hadronic uncertainties in the SI WIMP--nucleon cross section as a function of the Wilson coefficients 
for quark scalar and gluon effective operators. 


\section{Spin-independent Cross Section and Chiral Perturbation Theory}


In terms of the contributions from the dynamical degrees of freedom at the hadronic scale relevant for direct detection, the SI cross section for elastic Dirac WIMP scattering on a nucleon  ($N\in\{p,n\}$) has the form (cf.~\cite{Rajaraman:2011wf,HillSolon,Cirigliano:2012pq})\footnote{If the WIMP is a Majorana fermion, the right-hand side of~\eqref{SIeq} has to be multiplied by a factor of $4$. In~\eqref{SIeq}, contributions from tensor twist-2 operators are not present since we restrict ourselves to operators up to dimension 7.}
\begin{align}\label{SIeq}
\sigma _N^\text{SI}=& \frac{\mu _\chi^2}{\pi\,\Lambda^4}\bigg|\frac{m_N}{\Lambda}  \bigg(\sum\limits_{q = u,d,s}   C_{qq}^{SS}f_q^N -  12\pi \,C_{gg}^{S}\,f_Q^N  \bigg)
\notag\\
&+
\sum\limits_{q = u,d} C_{qq}^{VV}f_{V_q}^N \bigg|^2,
\end{align}
with $\mu_\chi = m_\chi m_N/(m_\chi + m_N)$ and scalar (vector) couplings $f_q^N$ ($f_{V_q}^N$).  For heavy quarks, the parameter $f_Q^N$ is induced by the gluon operator as discussed in~\cite{Shifman}. Accordingly, the Wilson coefficient $C_{gg}^S$ encodes matching corrections from integrating out $c$-, $b$-, and $t$-quarks as well as possible new heavier strongly interacting particles. The vector coefficients simply count
the valence quarks in a proton or a neutron, i.e.\ $f_{V_u}^p=f_{V_d}^n=2f_{V_d}^p=2f_{V_u}^n=2$, while the scalar couplings measure the contribution 
of the quark condensates to the mass of the nucleon
\begin{equation}
 \langle N|m_{q} \bar q q| N\rangle=f_q^N m_N.
\end{equation}
In the literature (see, e.g.~\cite{Ellis:2000ds,Ellis,Belanger:2008sj,Belanger:2013oya}) $f_u^N$ and $f_d^N$ are usually determined from the so-called strangeness content of the nucleon 
\begin{equation}
 y=\frac{2\langle N|\bar s s|N\rangle}{\langle N|\bar u u+\bar d d|N\rangle}
\end{equation}
and another quantity 
\begin{equation}
 z=\frac{\langle N|\bar u u-\bar s s|N\rangle}{\langle N|\bar d d-\bar s s|N\rangle}.
\end{equation}
The combination of $y$ and $z$ then permits the reconstruction of $f_u^N$ and $f_d^N$. $y$, in turn, is usually determined from $\sigma_{\pi N}$ based on $SU(3)$ ChPT~\cite{Borasoy}, an approach by itself afflicted with large uncertainties from the $SU(3)$ expansion. More crucially, it is not possible to attach a reliable error to the estimate $z\approx 1.49$ in~\cite{Cheng:1988im,Ellis:2000ds} commonly employed in the literature since it originates from leading-order fits to the baryon spectrum, whose inadequacy had already been demonstrated in~\cite{Gasser:1980sb,GL82}. Nevertheless, this value for $z$ has been widely used (see e.g.~\cite{Ellis,Belanger:2008sj,Belanger:2013oya}) without any attempt to quantify its inherent systematic uncertainty.

All these shortcomings can be avoided by using directly $SU(2)$ ChPT. 
The starting point is the chiral expansion of the nucleon mass in the presence of strong isospin violation~\cite{MS97,Muller:1999ww}
\begin{align}
\label{nucleon_mass}
m_N&=m_0-4c_1\mpii^2-\frac{e^2\Fpi^2}{2}(f_1\pm f_2+f_3)\\
&\pm2Bc_5(\md-\muu)-\frac{\ga^2\big(2M_{\pi^\pm}^3+\mpii^3\big)}{32\pi \Fpi^2}+\Order\big(\mpi^4\big),\notag
\end{align}
where the upper (lower) sign refers to proton (neutron), $B$ is related to the pion masses according to
\begin{align}
M_{\pi^\pm}^2&=B(\muu+\md)+2e^2\Fpi^2Z+\Order(m_q^2),\notag\\
 \mpii^2&=B(\muu+\md)+\Order(m_q^2),
\end{align}
$\Fpi$ denotes the pion decay constant, $e=\sqrt{4\pi\alpha}$ the electric charge, $\ga$ the axial coupling of the nucleon, 
 and $c_1$, $c_5$, $f_{1-3}$, $Z$ are low-energy constants, which encode short-distance effects.

The scalar couplings follow from~\eqref{nucleon_mass}
by means of
the Feynman--Hellmann theorem~\cite{Hellmann,Feynman39} 
\begin{equation}
\langle N | m_q\bar{q} q | N \rangle = m_q \frac{\partial m_N}{\partial m_q} \;\;\;\; {\text{with}}\;\;\; q \in \{u,d \},
\end{equation}
resulting in
\begin{align}
 f_u^N & = -\frac{2B}{m_N}
     \muu
  \Big[
     2c_1\pm c_5
    +\frac{9\ga^2\bar M_\pi}{128\pi\Fpi^2}\Big], \nonumber \\
f_d^N & = -\frac{2B}{m_N}     \md
  \Big[
     2c_1\mp c_5
       +\frac{9\ga^2\bar M_\pi}{128\pi\Fpi^2}\Big],   
\end{align}
where  
$\bar M_\pi=\big(2M_{\pi^\pm}+\mpii\big)/3$ denotes an average pion mass.
Next, we define $\sigma_{\pi N}$ as the average value of $1/2\,\langle N|(\muu+\md)(\bar u  u+\bar d d)|N\rangle$ between proton and neutron,\footnote{At this order in the chiral expansion the expressions for proton and neutron even coincide.}
which leads to the identification
\begin{equation}
\label{sigma_term}
\sigma_{\pi N}=-4c_1\mpii^2-\frac{9\ga^2\mpii^2\bar M_\pi}{64\pi\Fpi^2}+\Order(\mpi^4).
\end{equation}
This expression can be derived from~\eqref{nucleon_mass}, rewritten in terms of $\hat m=(\muu+\md)/2$ and the quark-mass difference, via another Feynman--Hellmann relation
\begin{equation}
 \sigma_{\pi N}=\frac{1}{2}\bigg(\hat m\frac{\partial \mpp}{\partial \hat m}+\hat m\frac{\partial \mn}{\partial \hat m}\bigg).
\end{equation}
In this way, we obtain the following result for the scalar couplings
\begin{align}
\label{ChPT_res}
 m_N  f_u^N &=\frac{\sigma_{\pi N}}{2}(1-\xi)\pm Bc_5\big(\md-\muu\big)   \Big(1-\frac{1}{\xi} \Big), \notag\\
 m_N   f_d^N &=\frac{\sigma_{\pi N}}{2}(1+\xi)\pm Bc_5\big(\md-\muu\big)    \Big(1+\frac{1}{\xi} \Big),\notag\\
\xi&=\frac{\md-\muu}{\md+\muu}=0.36\pm 0.04,
\end{align}
where again the upper (lower) sign refers to proton (neutron) and we used $\muu/\md=0.47\pm 0.04$ from~\cite{FLAG}.\footnote{In the isospin limit, this reduces to $m_Nf_u^N=m_Nf_d^N=\sigma_{\pi N}/2$, as expected~\cite{HillSolon}.}
Taking particle masses from~\cite{PDG} 
 and $B c_5(\md-\muu)=(-0.51\pm 0.08)\MeV$ according to the electromagnetic proton-neutron mass difference $(\mpp-\mn)^\text{em}=(0.76\pm 0.3)\MeV$ from~\cite{GL82},\footnote{Within uncertainties, this estimate for $c_5$, originating from an analysis of the Cottingham sum rule~\cite{Cottingham}, is consistent with a recent determination from a subtracted version of this sum rule with the subtraction constant estimated from nucleon polarizabilities~\cite{WL_Cottingham}, an extraction from $pn\to d \pi^0$~\cite{Filin}, and lattice calculations, see~\cite{Portelli} and references therein.} we find 
\begin{align}
\label{fq_res}
f_u^N&=\frac{\sigma_{\pi N}(1-\xi)}{2m_N}+\Delta f_u^N,\quad
f_d^N=\frac{\sigma_{\pi N}(1+\xi)}{2m_N}+\Delta f_d^N,\notag\\
  \Delta f_u^p
 &=(1.0\pm 0.2)\cdot 10^{-3},\quad 
   \Delta f_u^n
 =(-1.0\pm 0.2)\cdot 10^{-3},\notag\\
   \Delta f_d^p
 &=(-2.1\pm 0.4)\cdot 10^{-3},\quad
   \Delta f_d^n
 =(2.0\pm 0.4)\cdot 10^{-3}.
\end{align}
Expressing $c_1$ by means of~\eqref{sigma_term} can be understood as resumming higher chiral orders. We have verified this procedure explicitly at fourth order in the chiral expansion~\cite{Borasoy,FMS98,BL99}, with low-energy constants from~\cite{Bernard07} for a numerical analysis.
Our result shows that once $\sigma_{\pi N}$ is fixed, $f_u^N$ and $f_d^N$ can be inferred immediately, with both chiral expansion and isospin violation fully under control. This is crucial in order to accurately evaluate hadronic uncertainties in SI direct detection.

The importance of these findings for isospin-violating DM can be nicely illustrated by considering the difference between proton and neutron couplings
\begin{align}
\label{IV_us}
 f_u^p-f_u^n&=(1.9\pm 0.4)\cdot 10^{-3},\notag\\
f_d^p-f_d^n&=(-4.1\pm 0.7)\cdot 10^{-3}, 
\end{align}
where we used~\eqref{ChPT_res} directly, so that $\sigma_{\pi N}$ and $c_1$ drop out and the remaining uncertainty is generated by $c_5$ and $\muu/\md$. Comparing this result to the most recent estimate~\cite{Belanger:2013oya}
\begin{equation}
\label{IV_Micromegas}
 f_u^p-f_u^n=4.3\cdot 10^{-3},\quad
 f_d^p-f_d^n=-8.2\cdot 10^{-3},
\end{equation}
we see that the traditional approach overestimates isospin violation by a factor of $2$. As the difference between proton and neutron couplings is proportional to $c_5$, which measures the quark-mass contribution to the proton-neutron mass difference, this implies 
that the indirect reconstruction of this quantity by means of $y$ and $z$ fails by $100\%$.

A precise determination of the crucial $\sigma_{\pi N}$ is still an open issue. Ongoing efforts involve lattice QCD calculations at (nearly) physical values of the pion mass and refined phenomenological analyses. For a compilation of recent lattice results we refer to~\cite{Young,Kronfeld:2012uk,WalkerLoud,Belanger:2013oya} and references therein. The extraction of $\sigma_{\pi N}$ from $\pi N$ scattering 
requires an analytic continuation into the unphysical region~\cite{ChengDashen}, which is extremely sensitive to small shifts in the isoscalar amplitude, so that even isospin-breaking effects may become important.
On the experimental side, new information about threshold $\pi N$ scattering has become available over the last years thanks to accurate measurements in pionic atoms~\cite{Gotta:2008zza,Strauch:2010vu}. These results led
to a precision extraction of the $\pi N$ scattering lengths~\cite{piD,piDlong}, which are extremely valuable in stabilizing the analytic continuation.\footnote{In addition, these results for the scattering lengths nicely illustrate the sensitivity of the $\sigma$-term extraction to small changes in the isoscalar amplitude, as the isospin-breaking corrections~\cite{HKM,HKMlong} translated to $\sigma_{\pi N}$ according to~\cite{GLLS88} would lead to a shift of more than $5\MeV$.}
For these reasons, a systematic analysis of $\pi N$ scattering fully consistent with unitarity, analyticity, and crossing symmetric along the lines of~\cite{RS,RSSFF,HKMR}, respecting the new pionic-atom input, will help clarify the situation concerning the phenomenological determination of $\sigma_{\pi N}$~\cite{Gasser:1990ce,Pavan:2001wz,Alarcon:2011zs}.

Traditionally, the strangeness coupling $f_s^N$, or, equivalently, the strangeness content $y$, has been determined from $\sigma_{\pi N}$ based on $SU(3)$ ChPT~\cite{Borasoy}, incurring large uncertainties both from $\sigma_{\pi N}$ and the $SU(3)$ expansion. 
In view of recent lattice results, where contrary to the lightest quarks $m_s$ is close to its physical value,
a large strangeness content as sometimes inferred from $\sigma_{\pi N}$ becomes increasingly unlikely. 
In the following, we adopt the average from~\cite{WalkerLoud}
\begin{equation}
\label{fs_lattice}
f_s^N=0.043\pm 0.011,
\end{equation}
which takes into account the details of each lattice calculation in the averaging procedure.

Finally, the coupling for the heavy quarks is~\cite{Shifman}\footnote{For a discussion of $f_Q^N$ at higher orders in $\alpha_s$ we refer to~\cite{Kryjevski,Vecchi}.}
\begin{align}
 f_Q^N&=\frac{2}{27}\big(1-f_u^N-f_d^N-f_s^N\big).
\end{align}

\begin{figure}
\includegraphics[height=40ex]{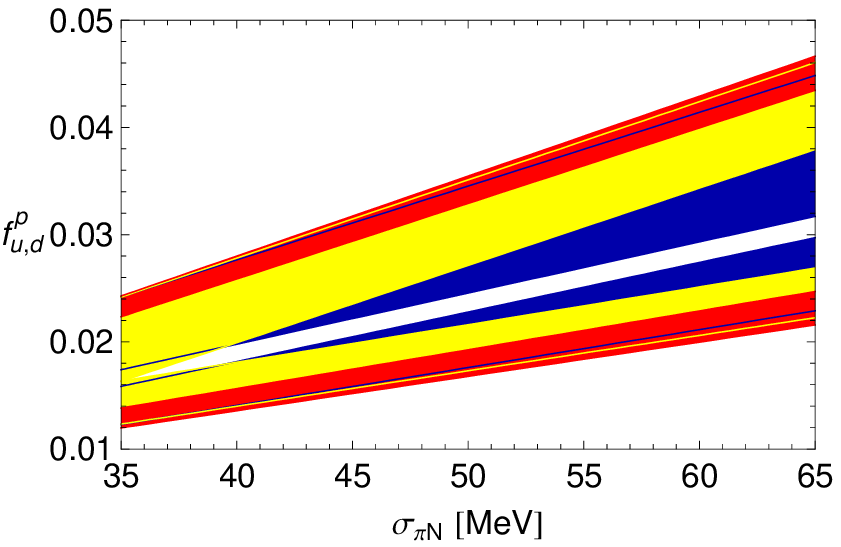}
\includegraphics[height=40ex]{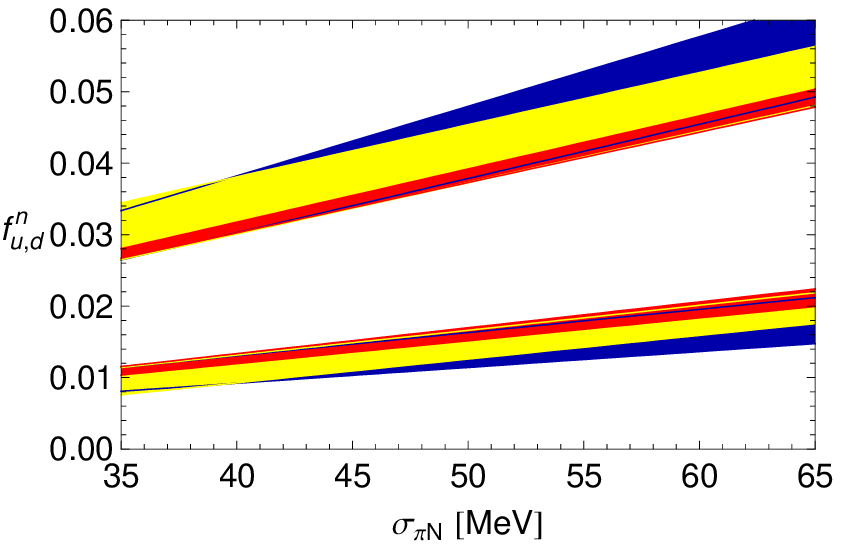}
\caption{Upper (lower panel): $f_u^p$ and $f_d^p$ ($f_u^n$ and $f_d^n$) as a function of $\sigma_{\pi N}$ according to~\eqref{fq_res} (red) compared to the traditional approach, with $y$ either derived from $\sigma_{\pi N}$ (yellow) or the lattice value~\eqref{fs_lattice} for $f_s^N$ (blue). In both plots, the upper (lower) bands refer to $d$- ($u$-)quark couplings.}
\label{fufdplots}
\end{figure}


\section{Numerical analysis}
\label{sec:numerics}


We first compare our results for the light-quark couplings to the traditional approach (see~\cite{Ellis,Belanger:2008sj}), as a function of $\sigma_{\pi N}$. Since in the latter case the $u$- and $d$- couplings are reconstructed from two strangeness-dependent quantities, we need to specify this input. We take $z=1.49$~\cite{Cheng:1988im,Ellis:2000ds}, and the strangeness content  $y$ either derived from the $SU(3)$ relation $y=1-\sigma_0/\sigma_{\pi N}$, with $\sigma_0=(36\pm 7)\MeV$~\cite{Borasoy}, 
or fixed from the lattice value~\eqref{fs_lattice} for $f_s^N$ via
\begin{equation}
y= \frac{m_Nf_s^N}{\sigma_{\pi N}} \frac{2\hat m}{m_s}.
\end{equation}
Without resorting to higher-order calculations for $z$, as usually done in the literature, it is impossible to provide a reliable uncertainty estimate for this quantity. Based on general expectations of the convergence of the $SU(3)$ expansion, we simply take a $30\%$ error. In fact, the large shift 
between the leading-order value $\sigma_0\simeq 26\MeV$, as extracted from hadron masses in analogy to~\cite{Cheng:1988im}, to $\sigma_0\simeq (36\pm 7)\MeV$ due to higher chiral orders indicates that the inherent uncertainty may be even larger.

As shown in Fig.~\ref{fufdplots}, for both determinations of $y$ we observe a moderate shift of the central value or a change in slope, compared to our approach. More importantly, the band for a given value of $\sigma_{\pi N}$ shrinks drastically.
This shows that  one can take proper advantage of a precise determination of $\sigma_{\pi N}$, with accurate error estimates, only within our framework, since otherwise the need for strangeness input thwarts the transition to the two-flavor scalar couplings. Due to the arbitrariness in estimating the uncertainties of this strangeness input, especially of $z$, our approach is the only way to achieve reliable error estimates.

\begin{figure}
\includegraphics[height=40ex]{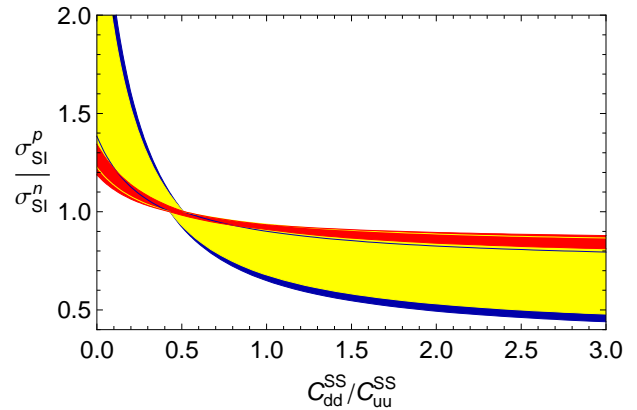}
\caption{Ratio of proton and neutron cross section for $C^{SS}_{ss}=C_{gg}^S=C_{qq}^{VV}=0$ and $\sigma_{\pi N}=50\,\MeV$. This illustrates
the maximally possible isospin violation induced by scalar operators. Color coding as in Fig.~\ref{fufdplots}.}
\label{pn}
\end{figure}

Constraining the Wilson coefficients $C_{gg}^S$ and $C_{qq}^{SS}$, see~\eqref{Lagr} and \eqref{SIeq}, allows one to gain information about DM-Higgs operators from direct detection~\cite{LopezHonorez:2012kv}, by proper renormalization group evolution, matching corrections~\cite{Shifman}, and mixing~\cite{Frandsen:2012db}, from the low-energy hadronic scale up to the scale $\Lambda$ of New Physics~\cite{CEP}.
The impact of our results compared to the traditional approach becomes most pronounced in the context of isospin violation.
In the absence of vector operators, the ratio $C^{SS}_{dd}/C^{SS}_{uu}$ is the quantity responsible for isospin-violating effects. 
Models with isospin violation in the scalar sector have been considered e.g.\ in~\cite{Giuliani:2005my,Ellis,Gao:2011ka,Okada:2013cba,Belanger:2013tla}. It has been argued that even in the Constrained MSSM isospin violation could be large enough to be detected in experiment~\cite{Ellis}.
In Fig.~\ref{pn} we show the ratio of SI WIMP--proton and WIMP--neutron cross sections as a function of $C^{SS}_{dd}/C^{SS}_{uu}$, assuming that all other Wilson coefficients are zero. Again, we see that using our approach the uncertainties reduce drastically, while hinting at smaller isospin-violating effects than expected  before, see~\eqref{IV_us} and~\eqref{IV_Micromegas}.

\section{Conclusions}

In this article we have presented a novel approach to determine the proton and neutron scalar couplings $f_u^{p,n}$ and $f_d^{p,n}$, which are key input quantities for direct DM searches.
Our central results are the expressions given in~\eqref{ChPT_res} and~\eqref{fq_res} based on $SU(2)$ ChPT.
We have provided values for these coefficients, as a function of the pion--nucleon $\sigma$-term,
without any reference to an $SU(3)$ expansion and consistently incorporating isospin-violating effects. 
Thus removing an additional source of theoretical uncertainty that had so far been overlooked in the literature, our results
permit an honest assessment of hadronic uncertainties in DM detection without uncontrolled approximations.

\section*{Acknowledgments}

We thank H.~Leutwyler for helpful discussions and B.~Kubis and U.-G.~Mei{\ss}ner for comments on the manuscript. Support by the Swiss National Science Foundation (SNF) and by the ``Innovations- und Kooperationsprojekt C-13" of the Schweizerische Universit{\"a}tskonferenz SUK/CRUS is gratefully acknowledged.

\bibliography{BIB_AMM_v2}

\begin{thebibliography}{65}
\expandafter\ifx\csname natexlab\endcsname\relax\def\natexlab#1{#1}\fi
\expandafter\ifx\csname bibnamefont\endcsname\relax
  \def\bibnamefont#1{#1}\fi
\expandafter\ifx\csname bibfnamefont\endcsname\relax
  \def\bibfnamefont#1{#1}\fi
\expandafter\ifx\csname citenamefont\endcsname\relax
  \def\citenamefont#1{#1}\fi
\expandafter\ifx\csname url\endcsname\relax
  \def\url#1{\texttt{#1}}\fi
\expandafter\ifx\csname urlprefix\endcsname\relax\def\urlprefix{URL }\fi
\providecommand{\bibinfo}[2]{#2}
\providecommand{\eprint}[2][]{\url{#2}}

\bibitem[{\citenamefont{Bernabei et~al.}(2008)}]{Bernabei:2008yi}
\bibinfo{author}{\bibfnamefont{R.}~\bibnamefont{Bernabei}} \bibnamefont{et~al.}
  (\bibinfo{collaboration}{DAMA Collaboration}), \bibinfo{journal}{Eur.Phys.J.}
  \textbf{\bibinfo{volume}{C56}}, \bibinfo{pages}{333} (\bibinfo{year}{2008}),
  \eprint{0804.2741}.

\bibitem[{\citenamefont{Aalseth et~al.}(2011)}]{Aalseth:2011wp}
\bibinfo{author}{\bibfnamefont{C.}~\bibnamefont{Aalseth}} \bibnamefont{et~al.},
  \bibinfo{journal}{Phys.Rev.Lett.} \textbf{\bibinfo{volume}{107}},
  \bibinfo{pages}{141301} (\bibinfo{year}{2011}), \eprint{1106.0650}.

\bibitem[{\citenamefont{Angloher et~al.}(2012)}]{Angloher:2011uu}
\bibinfo{author}{\bibfnamefont{G.}~\bibnamefont{Angloher}}
  \bibnamefont{et~al.}, \bibinfo{journal}{Eur.Phys.J.}
  \textbf{\bibinfo{volume}{C72}}, \bibinfo{pages}{1971} (\bibinfo{year}{2012}),
  \eprint{1109.0702}.

\bibitem[{\citenamefont{Agnese et~al.}(2013)}]{Agnese:2013rvf}
\bibinfo{author}{\bibfnamefont{R.}~\bibnamefont{Agnese}} \bibnamefont{et~al.}
  (\bibinfo{collaboration}{CDMS Collaboration}),
  \bibinfo{journal}{Phys.Rev.Lett.}  (\bibinfo{year}{2013}),
  \eprint{1304.4279}.

\bibitem[{\citenamefont{Aprile et~al.}(2011)}]{Aprile:2011hi}
\bibinfo{author}{\bibfnamefont{E.}~\bibnamefont{Aprile}} \bibnamefont{et~al.}
  (\bibinfo{collaboration}{XENON100 Collaboration}),
  \bibinfo{journal}{Phys.Rev.Lett.} \textbf{\bibinfo{volume}{107}},
  \bibinfo{pages}{131302} (\bibinfo{year}{2011}), \eprint{1104.2549}.

\bibitem[{\citenamefont{Angle et~al.}(2011)}]{Angle:2011th}
\bibinfo{author}{\bibfnamefont{J.}~\bibnamefont{Angle}} \bibnamefont{et~al.}
  (\bibinfo{collaboration}{XENON10 Collaboration}),
  \bibinfo{journal}{Phys.Rev.Lett.} \textbf{\bibinfo{volume}{107}},
  \bibinfo{pages}{051301} (\bibinfo{year}{2011}), \eprint{1104.3088}.

\bibitem[{\citenamefont{Akerib et~al.}(2013)}]{Akerib:2013tjd}
\bibinfo{author}{\bibfnamefont{D.}~\bibnamefont{Akerib}} \bibnamefont{et~al.}
  (\bibinfo{collaboration}{LUX Collaboration}) (\bibinfo{year}{2013}),
  \eprint{1310.8214}.

\bibitem[{\citenamefont{Cushman et~al.}(2013)}]{Snowmass}
\bibinfo{author}{\bibfnamefont{P.}~\bibnamefont{Cushman}} \bibnamefont{et~al.}
  (\bibinfo{year}{2013}), \eprint{1310.8327}.

\bibitem[{\citenamefont{Cheng}(1989)}]{Cheng:1988im}
\bibinfo{author}{\bibfnamefont{H.-Y.} \bibnamefont{Cheng}},
  \bibinfo{journal}{Phys.Lett.} \textbf{\bibinfo{volume}{B219}},
  \bibinfo{pages}{347} (\bibinfo{year}{1989}).

\bibitem[{\citenamefont{Bottino et~al.}(2000)\citenamefont{Bottino, Donato,
  Fornengo, and Scopel}}]{Bottino:1999ei}
\bibinfo{author}{\bibfnamefont{A.}~\bibnamefont{Bottino}},
  \bibinfo{author}{\bibfnamefont{F.}~\bibnamefont{Donato}},
  \bibinfo{author}{\bibfnamefont{N.}~\bibnamefont{Fornengo}}, \bibnamefont{and}
  \bibinfo{author}{\bibfnamefont{S.}~\bibnamefont{Scopel}},
  \bibinfo{journal}{Astropart.Phys.} \textbf{\bibinfo{volume}{13}},
  \bibinfo{pages}{215} (\bibinfo{year}{2000}), \eprint{hep-ph/9909228}.

\bibitem[{\citenamefont{Bottino et~al.}(2002)\citenamefont{Bottino, Donato,
  Fornengo, and Scopel}}]{Bottino:2001dj}
\bibinfo{author}{\bibfnamefont{A.}~\bibnamefont{Bottino}},
  \bibinfo{author}{\bibfnamefont{F.}~\bibnamefont{Donato}},
  \bibinfo{author}{\bibfnamefont{N.}~\bibnamefont{Fornengo}}, \bibnamefont{and}
  \bibinfo{author}{\bibfnamefont{S.}~\bibnamefont{Scopel}},
  \bibinfo{journal}{Astropart.Phys.} \textbf{\bibinfo{volume}{18}},
  \bibinfo{pages}{205} (\bibinfo{year}{2002}), \eprint{hep-ph/0111229}.

\bibitem[{\citenamefont{Ellis et~al.}(2008)\citenamefont{Ellis, Olive, and
  Savage}}]{Ellis}
\bibinfo{author}{\bibfnamefont{J.~R.} \bibnamefont{Ellis}},
  \bibinfo{author}{\bibfnamefont{K.~A.} \bibnamefont{Olive}}, \bibnamefont{and}
  \bibinfo{author}{\bibfnamefont{C.}~\bibnamefont{Savage}},
  \bibinfo{journal}{Phys.Rev.} \textbf{\bibinfo{volume}{D77}},
  \bibinfo{pages}{065026} (\bibinfo{year}{2008}), \eprint{0801.3656}.

\bibitem[{\citenamefont{Hill and Solon}(2012)}]{HillSolon}
\bibinfo{author}{\bibfnamefont{R.~J.} \bibnamefont{Hill}} \bibnamefont{and}
  \bibinfo{author}{\bibfnamefont{M.~P.} \bibnamefont{Solon}},
  \bibinfo{journal}{Phys.Lett.} \textbf{\bibinfo{volume}{B707}},
  \bibinfo{pages}{539} (\bibinfo{year}{2012}), \eprint{1111.0016}.

\bibitem[{\citenamefont{Kurylov and Kamionkowski}(2004)}]{Kurylov:2003ra}
\bibinfo{author}{\bibfnamefont{A.}~\bibnamefont{Kurylov}} \bibnamefont{and}
  \bibinfo{author}{\bibfnamefont{M.}~\bibnamefont{Kamionkowski}},
  \bibinfo{journal}{Phys.Rev.} \textbf{\bibinfo{volume}{D69}},
  \bibinfo{pages}{063503} (\bibinfo{year}{2004}), \eprint{hep-ph/0307185}.

\bibitem[{\citenamefont{Giuliani}(2005)}]{Giuliani:2005my}
\bibinfo{author}{\bibfnamefont{F.}~\bibnamefont{Giuliani}},
  \bibinfo{journal}{Phys.Rev.Lett.} \textbf{\bibinfo{volume}{95}},
  \bibinfo{pages}{101301} (\bibinfo{year}{2005}), \eprint{hep-ph/0504157}.

\bibitem[{\citenamefont{Chang et~al.}(2010)\citenamefont{Chang, Liu, Pierce,
  Weiner, and Yavin}}]{Chang:2010yk}
\bibinfo{author}{\bibfnamefont{S.}~\bibnamefont{Chang}},
  \bibinfo{author}{\bibfnamefont{J.}~\bibnamefont{Liu}},
  \bibinfo{author}{\bibfnamefont{A.}~\bibnamefont{Pierce}},
  \bibinfo{author}{\bibfnamefont{N.}~\bibnamefont{Weiner}}, \bibnamefont{and}
  \bibinfo{author}{\bibfnamefont{I.}~\bibnamefont{Yavin}},
  \bibinfo{journal}{JCAP} \textbf{\bibinfo{volume}{1008}}, \bibinfo{pages}{018}
  (\bibinfo{year}{2010}), \eprint{1004.0697}.

\bibitem[{\citenamefont{Feng et~al.}(2011)\citenamefont{Feng, Kumar, Marfatia,
  and Sanford}}]{Feng:2011vu}
\bibinfo{author}{\bibfnamefont{J.~L.} \bibnamefont{Feng}},
  \bibinfo{author}{\bibfnamefont{J.}~\bibnamefont{Kumar}},
  \bibinfo{author}{\bibfnamefont{D.}~\bibnamefont{Marfatia}}, \bibnamefont{and}
  \bibinfo{author}{\bibfnamefont{D.}~\bibnamefont{Sanford}},
  \bibinfo{journal}{Phys.Lett.} \textbf{\bibinfo{volume}{B703}},
  \bibinfo{pages}{124} (\bibinfo{year}{2011}), \eprint{1102.4331}.

\bibitem[{\citenamefont{Cirigliano et~al.}(2012)\citenamefont{Cirigliano,
  Graesser, and Ovanesyan}}]{Cirigliano:2012pq}
\bibinfo{author}{\bibfnamefont{V.}~\bibnamefont{Cirigliano}},
  \bibinfo{author}{\bibfnamefont{M.~L.} \bibnamefont{Graesser}},
  \bibnamefont{and}
  \bibinfo{author}{\bibfnamefont{G.}~\bibnamefont{Ovanesyan}},
  \bibinfo{journal}{JHEP} \textbf{\bibinfo{volume}{1210}}, \bibinfo{pages}{025}
  (\bibinfo{year}{2012}), \eprint{1205.2695}.

\bibitem[{\citenamefont{Cirigliano et~al.}(2013)\citenamefont{Cirigliano,
  Graesser, Ovanesyan, and Shoemaker}}]{Cirigliano13}
\bibinfo{author}{\bibfnamefont{V.}~\bibnamefont{Cirigliano}},
  \bibinfo{author}{\bibfnamefont{M.~L.} \bibnamefont{Graesser}},
  \bibinfo{author}{\bibfnamefont{G.}~\bibnamefont{Ovanesyan}},
  \bibnamefont{and} \bibinfo{author}{\bibfnamefont{I.~M.}
  \bibnamefont{Shoemaker}} (\bibinfo{year}{2013}), \eprint{1311.5886}.

\bibitem[{\citenamefont{Rajaraman et~al.}(2011)\citenamefont{Rajaraman,
  Shepherd, Tait, and Wijangco}}]{Rajaraman:2011wf}
\bibinfo{author}{\bibfnamefont{A.}~\bibnamefont{Rajaraman}},
  \bibinfo{author}{\bibfnamefont{W.}~\bibnamefont{Shepherd}},
  \bibinfo{author}{\bibfnamefont{T.~M.} \bibnamefont{Tait}}, \bibnamefont{and}
  \bibinfo{author}{\bibfnamefont{A.~M.} \bibnamefont{Wijangco}},
  \bibinfo{journal}{Phys.Rev.} \textbf{\bibinfo{volume}{D84}},
  \bibinfo{pages}{095013} (\bibinfo{year}{2011}), \eprint{1108.1196}.

\bibitem[{\citenamefont{Shifman et~al.}(1978)\citenamefont{Shifman, Vainshtein,
  and Zakharov}}]{Shifman}
\bibinfo{author}{\bibfnamefont{M.~A.} \bibnamefont{Shifman}},
  \bibinfo{author}{\bibfnamefont{A.}~\bibnamefont{Vainshtein}},
  \bibnamefont{and} \bibinfo{author}{\bibfnamefont{V.~I.}
  \bibnamefont{Zakharov}}, \bibinfo{journal}{Phys.Lett.}
  \textbf{\bibinfo{volume}{B78}}, \bibinfo{pages}{443} (\bibinfo{year}{1978}).

\bibitem[{\citenamefont{Ellis et~al.}(2000)\citenamefont{Ellis, Ferstl, and
  Olive}}]{Ellis:2000ds}
\bibinfo{author}{\bibfnamefont{J.~R.} \bibnamefont{Ellis}},
  \bibinfo{author}{\bibfnamefont{A.}~\bibnamefont{Ferstl}}, \bibnamefont{and}
  \bibinfo{author}{\bibfnamefont{K.~A.} \bibnamefont{Olive}},
  \bibinfo{journal}{Phys.Lett.} \textbf{\bibinfo{volume}{B481}},
  \bibinfo{pages}{304} (\bibinfo{year}{2000}), \eprint{hep-ph/0001005}.

\bibitem[{\citenamefont{B{\'e}langer et~al.}(2009)\citenamefont{B{\'e}langer,
  Boudjema, Pukhov, and Semenov}}]{Belanger:2008sj}
\bibinfo{author}{\bibfnamefont{G.}~\bibnamefont{B{\'e}langer}},
  \bibinfo{author}{\bibfnamefont{F.}~\bibnamefont{Boudjema}},
  \bibinfo{author}{\bibfnamefont{A.}~\bibnamefont{Pukhov}}, \bibnamefont{and}
  \bibinfo{author}{\bibfnamefont{A.}~\bibnamefont{Semenov}},
  \bibinfo{journal}{Comput.Phys.Commun.} \textbf{\bibinfo{volume}{180}},
  \bibinfo{pages}{747} (\bibinfo{year}{2009}), \eprint{0803.2360}.

\bibitem[{\citenamefont{B{\'e}langer
  et~al.}(2013{\natexlab{a}})\citenamefont{B{\'e}langer, Boudjema, Pukhov, and
  Semenov}}]{Belanger:2013oya}
\bibinfo{author}{\bibfnamefont{G.}~\bibnamefont{B{\'e}langer}},
  \bibinfo{author}{\bibfnamefont{F.}~\bibnamefont{Boudjema}},
  \bibinfo{author}{\bibfnamefont{A.}~\bibnamefont{Pukhov}}, \bibnamefont{and}
  \bibinfo{author}{\bibfnamefont{A.}~\bibnamefont{Semenov}}
  (\bibinfo{year}{2013}{\natexlab{a}}), \eprint{1305.0237}.

\bibitem[{\citenamefont{Borasoy and Mei{\ss}ner}(1997)}]{Borasoy}
\bibinfo{author}{\bibfnamefont{B.}~\bibnamefont{Borasoy}} \bibnamefont{and}
  \bibinfo{author}{\bibfnamefont{U.-G.} \bibnamefont{Mei{\ss}ner}},
  \bibinfo{journal}{Annals Phys.} \textbf{\bibinfo{volume}{254}},
  \bibinfo{pages}{192} (\bibinfo{year}{1997}), \eprint{hep-ph/9607432}.

\bibitem[{\citenamefont{Gasser}(1981)}]{Gasser:1980sb}
\bibinfo{author}{\bibfnamefont{J.}~\bibnamefont{Gasser}},
  \bibinfo{journal}{Annals Phys.} \textbf{\bibinfo{volume}{136}},
  \bibinfo{pages}{62} (\bibinfo{year}{1981}).

\bibitem[{\citenamefont{Gasser and Leutwyler}(1982)}]{GL82}
\bibinfo{author}{\bibfnamefont{J.}~\bibnamefont{Gasser}} \bibnamefont{and}
  \bibinfo{author}{\bibfnamefont{H.}~\bibnamefont{Leutwyler}},
  \bibinfo{journal}{Phys.Rept.} \textbf{\bibinfo{volume}{87}},
  \bibinfo{pages}{77} (\bibinfo{year}{1982}).

\bibitem[{\citenamefont{Mei{\ss}ner and Steininger}(1998)}]{MS97}
\bibinfo{author}{\bibfnamefont{U.-G.} \bibnamefont{Mei{\ss}ner}}
  \bibnamefont{and}
  \bibinfo{author}{\bibfnamefont{S.}~\bibnamefont{Steininger}},
  \bibinfo{journal}{Phys.Lett.} \textbf{\bibinfo{volume}{B419}},
  \bibinfo{pages}{403} (\bibinfo{year}{1998}), \eprint{hep-ph/9709453}.

\bibitem[{\citenamefont{M{\"u}ller and Mei{\ss}ner}(1999)}]{Muller:1999ww}
\bibinfo{author}{\bibfnamefont{G.}~\bibnamefont{M{\"u}ller}} \bibnamefont{and}
  \bibinfo{author}{\bibfnamefont{U.-G.} \bibnamefont{Mei{\ss}ner}},
  \bibinfo{journal}{Nucl.Phys.} \textbf{\bibinfo{volume}{B556}},
  \bibinfo{pages}{265} (\bibinfo{year}{1999}), \eprint{hep-ph/9903375}.

\bibitem[{\citenamefont{Hellmann}(1937)}]{Hellmann}
\bibinfo{author}{\bibfnamefont{H.}~\bibnamefont{Hellmann}},
  \emph{\bibinfo{title}{{Einf\"uhrung in die Quantenchemie}}}
  (\bibinfo{publisher}{Franz Deuticke, Leipzig}, \bibinfo{year}{1937}).

\bibitem[{\citenamefont{Feynman}(1939)}]{Feynman39}
\bibinfo{author}{\bibfnamefont{R.}~\bibnamefont{Feynman}},
  \bibinfo{journal}{Phys.Rev.} \textbf{\bibinfo{volume}{56}},
  \bibinfo{pages}{340} (\bibinfo{year}{1939}).

\bibitem[{\citenamefont{Colangelo et~al.}(2011)}]{FLAG}
\bibinfo{author}{\bibfnamefont{G.}~\bibnamefont{Colangelo}}
  \bibnamefont{et~al.}, \bibinfo{journal}{Eur.Phys.J.}
  \textbf{\bibinfo{volume}{C71}}, \bibinfo{pages}{1695} (\bibinfo{year}{2011}),
  \eprint{1011.4408}.

\bibitem[{\citenamefont{Beringer et~al.}(2012)}]{PDG}
\bibinfo{author}{\bibfnamefont{J.}~\bibnamefont{Beringer}} \bibnamefont{et~al.}
  (\bibinfo{collaboration}{Particle Data Group}), \bibinfo{journal}{Phys.Rev.}
  \textbf{\bibinfo{volume}{D86}}, \bibinfo{pages}{010001}
  (\bibinfo{year}{2012}).

\bibitem[{\citenamefont{Cottingham}(1963)}]{Cottingham}
\bibinfo{author}{\bibfnamefont{W.}~\bibnamefont{Cottingham}},
  \bibinfo{journal}{Annals Phys.} \textbf{\bibinfo{volume}{25}},
  \bibinfo{pages}{424} (\bibinfo{year}{1963}).

\bibitem[{\citenamefont{Walker-Loud et~al.}(2012)\citenamefont{Walker-Loud,
  Carlson, and Miller}}]{WL_Cottingham}
\bibinfo{author}{\bibfnamefont{A.}~\bibnamefont{Walker-Loud}},
  \bibinfo{author}{\bibfnamefont{C.~E.} \bibnamefont{Carlson}},
  \bibnamefont{and} \bibinfo{author}{\bibfnamefont{G.~A.}
  \bibnamefont{Miller}}, \bibinfo{journal}{Phys.Rev.Lett.}
  \textbf{\bibinfo{volume}{108}}, \bibinfo{pages}{232301}
  (\bibinfo{year}{2012}), \eprint{1203.0254}.

\bibitem[{\citenamefont{Filin et~al.}(2009)}]{Filin}
\bibinfo{author}{\bibfnamefont{A.}~\bibnamefont{Filin}} \bibnamefont{et~al.},
  \bibinfo{journal}{Phys.Lett.} \textbf{\bibinfo{volume}{B681}},
  \bibinfo{pages}{423} (\bibinfo{year}{2009}), \eprint{0907.4671}.

\bibitem[{\citenamefont{Portelli}(2013)}]{Portelli}
\bibinfo{author}{\bibfnamefont{A.}~\bibnamefont{Portelli}},
  \bibinfo{journal}{PoS} \textbf{\bibinfo{volume}{KAON13}},
  \bibinfo{pages}{023} (\bibinfo{year}{2013}), \eprint{1307.6056}.

\bibitem[{\citenamefont{Steininger et~al.}(1998)\citenamefont{Steininger,
  Mei{\ss}ner, and Fettes}}]{FMS98}
\bibinfo{author}{\bibfnamefont{S.}~\bibnamefont{Steininger}},
  \bibinfo{author}{\bibfnamefont{U.-G.} \bibnamefont{Mei{\ss}ner}},
  \bibnamefont{and} \bibinfo{author}{\bibfnamefont{N.}~\bibnamefont{Fettes}},
  \bibinfo{journal}{JHEP} \textbf{\bibinfo{volume}{9809}}, \bibinfo{pages}{008}
  (\bibinfo{year}{1998}), \eprint{hep-ph/9808280}.

\bibitem[{\citenamefont{Becher and Leutwyler}(1999)}]{BL99}
\bibinfo{author}{\bibfnamefont{T.}~\bibnamefont{Becher}} \bibnamefont{and}
  \bibinfo{author}{\bibfnamefont{H.}~\bibnamefont{Leutwyler}},
  \bibinfo{journal}{Eur.Phys.J.} \textbf{\bibinfo{volume}{C9}},
  \bibinfo{pages}{643} (\bibinfo{year}{1999}), \eprint{hep-ph/9901384}.

\bibitem[{\citenamefont{Bernard}(2008)}]{Bernard07}
\bibinfo{author}{\bibfnamefont{V.}~\bibnamefont{Bernard}},
  \bibinfo{journal}{Prog.Part.Nucl.Phys.} \textbf{\bibinfo{volume}{60}},
  \bibinfo{pages}{82} (\bibinfo{year}{2008}), \eprint{0706.0312}.

\bibitem[{\citenamefont{Young}(2012)}]{Young}
\bibinfo{author}{\bibfnamefont{R.}~\bibnamefont{Young}}, \bibinfo{journal}{PoS}
  \textbf{\bibinfo{volume}{LATTICE2012}}, \bibinfo{pages}{014}
  (\bibinfo{year}{2012}), \eprint{1301.1765}.

\bibitem[{\citenamefont{Kronfeld}(2012)}]{Kronfeld:2012uk}
\bibinfo{author}{\bibfnamefont{A.~S.} \bibnamefont{Kronfeld}},
  \bibinfo{journal}{Ann.Rev.Nucl.Part.Sci.} \textbf{\bibinfo{volume}{62}},
  \bibinfo{pages}{265} (\bibinfo{year}{2012}), \eprint{1203.1204}.

\bibitem[{\citenamefont{Junnarkar and Walker-Loud}(2013)}]{WalkerLoud}
\bibinfo{author}{\bibfnamefont{P.}~\bibnamefont{Junnarkar}} \bibnamefont{and}
  \bibinfo{author}{\bibfnamefont{A.}~\bibnamefont{Walker-Loud}},
  \bibinfo{journal}{Phys.Rev.} \textbf{\bibinfo{volume}{D87}},
  \bibinfo{pages}{114510} (\bibinfo{year}{2013}), \eprint{1301.1114}.

\bibitem[{\citenamefont{Cheng and Dashen}(1971)}]{ChengDashen}
\bibinfo{author}{\bibfnamefont{T.}~\bibnamefont{Cheng}} \bibnamefont{and}
  \bibinfo{author}{\bibfnamefont{R.~F.} \bibnamefont{Dashen}},
  \bibinfo{journal}{Phys.Rev.Lett.} \textbf{\bibinfo{volume}{26}},
  \bibinfo{pages}{594} (\bibinfo{year}{1971}).

\bibitem[{\citenamefont{Gotta et~al.}(2008)}]{Gotta:2008zza}
\bibinfo{author}{\bibfnamefont{D.}~\bibnamefont{Gotta}} \bibnamefont{et~al.},
  \bibinfo{journal}{Lect.Notes Phys.} \textbf{\bibinfo{volume}{745}},
  \bibinfo{pages}{165} (\bibinfo{year}{2008}).

\bibitem[{\citenamefont{Strauch et~al.}(2011)}]{Strauch:2010vu}
\bibinfo{author}{\bibfnamefont{T.}~\bibnamefont{Strauch}} \bibnamefont{et~al.},
  \bibinfo{journal}{Eur.Phys.J.} \textbf{\bibinfo{volume}{A47}},
  \bibinfo{pages}{88} (\bibinfo{year}{2011}), \eprint{1011.2415}.

\bibitem[{\citenamefont{Baru et~al.}(2011{\natexlab{a}})}]{piD}
\bibinfo{author}{\bibfnamefont{V.}~\bibnamefont{Baru}} \bibnamefont{et~al.},
  \bibinfo{journal}{Phys.Lett.} \textbf{\bibinfo{volume}{B694}},
  \bibinfo{pages}{473} (\bibinfo{year}{2011}{\natexlab{a}}),
  \eprint{1003.4444}.

\bibitem[{\citenamefont{Baru et~al.}(2011{\natexlab{b}})}]{piDlong}
\bibinfo{author}{\bibfnamefont{V.}~\bibnamefont{Baru}} \bibnamefont{et~al.},
  \bibinfo{journal}{Nucl.Phys.} \textbf{\bibinfo{volume}{A872}},
  \bibinfo{pages}{69} (\bibinfo{year}{2011}{\natexlab{b}}), \eprint{1107.5509}.

\bibitem[{\citenamefont{Hoferichter et~al.}(2009)\citenamefont{Hoferichter,
  Kubis, and Mei{\ss}ner}}]{HKM}
\bibinfo{author}{\bibfnamefont{M.}~\bibnamefont{Hoferichter}},
  \bibinfo{author}{\bibfnamefont{B.}~\bibnamefont{Kubis}}, \bibnamefont{and}
  \bibinfo{author}{\bibfnamefont{U.-G.} \bibnamefont{Mei{\ss}ner}},
  \bibinfo{journal}{Phys.Lett.} \textbf{\bibinfo{volume}{B678}},
  \bibinfo{pages}{65} (\bibinfo{year}{2009}), \eprint{0903.3890}.

\bibitem[{\citenamefont{Hoferichter et~al.}(2010)\citenamefont{Hoferichter,
  Kubis, and Mei{\ss}ner}}]{HKMlong}
\bibinfo{author}{\bibfnamefont{M.}~\bibnamefont{Hoferichter}},
  \bibinfo{author}{\bibfnamefont{B.}~\bibnamefont{Kubis}}, \bibnamefont{and}
  \bibinfo{author}{\bibfnamefont{U.-G.} \bibnamefont{Mei{\ss}ner}},
  \bibinfo{journal}{Nucl.Phys.} \textbf{\bibinfo{volume}{A833}},
  \bibinfo{pages}{18} (\bibinfo{year}{2010}), \eprint{0909.4390}.

\bibitem[{\citenamefont{Gasser et~al.}(1988)\citenamefont{Gasser, Leutwyler,
  Locher, and Sainio}}]{GLLS88}
\bibinfo{author}{\bibfnamefont{J.}~\bibnamefont{Gasser}},
  \bibinfo{author}{\bibfnamefont{H.}~\bibnamefont{Leutwyler}},
  \bibinfo{author}{\bibfnamefont{M.}~\bibnamefont{Locher}}, \bibnamefont{and}
  \bibinfo{author}{\bibfnamefont{M.}~\bibnamefont{Sainio}},
  \bibinfo{journal}{Phys.Lett.} \textbf{\bibinfo{volume}{B213}},
  \bibinfo{pages}{85} (\bibinfo{year}{1988}).

\bibitem[{\citenamefont{Ditsche et~al.}(2012)\citenamefont{Ditsche,
  Hoferichter, Kubis, and Mei{\ss}ner}}]{RS}
\bibinfo{author}{\bibfnamefont{C.}~\bibnamefont{Ditsche}},
  \bibinfo{author}{\bibfnamefont{M.}~\bibnamefont{Hoferichter}},
  \bibinfo{author}{\bibfnamefont{B.}~\bibnamefont{Kubis}}, \bibnamefont{and}
  \bibinfo{author}{\bibfnamefont{U.-G.} \bibnamefont{Mei{\ss}ner}},
  \bibinfo{journal}{JHEP} \textbf{\bibinfo{volume}{1206}}, \bibinfo{pages}{043}
  (\bibinfo{year}{2012}), \eprint{1203.4758}.

\bibitem[{\citenamefont{Hoferichter et~al.}(2012)\citenamefont{Hoferichter,
  Ditsche, Kubis, and Mei{\ss}ner}}]{RSSFF}
\bibinfo{author}{\bibfnamefont{M.}~\bibnamefont{Hoferichter}},
  \bibinfo{author}{\bibfnamefont{C.}~\bibnamefont{Ditsche}},
  \bibinfo{author}{\bibfnamefont{B.}~\bibnamefont{Kubis}}, \bibnamefont{and}
  \bibinfo{author}{\bibfnamefont{U.-G.} \bibnamefont{Mei{\ss}ner}},
  \bibinfo{journal}{JHEP} \textbf{\bibinfo{volume}{1206}}, \bibinfo{pages}{063}
  (\bibinfo{year}{2012}), \eprint{1204.6251}.

\bibitem[{\citenamefont{Hoferichter et~al.}(in
  preparation)\citenamefont{Hoferichter, Ruiz~de Elvira, Kubis, and
  Mei{\ss}ner}}]{HKMR}
\bibinfo{author}{\bibfnamefont{M.}~\bibnamefont{Hoferichter}},
  \bibinfo{author}{\bibfnamefont{J.}~\bibnamefont{Ruiz~de Elvira}},
  \bibinfo{author}{\bibfnamefont{B.}~\bibnamefont{Kubis}}, \bibnamefont{and}
  \bibinfo{author}{\bibfnamefont{U.-G.} \bibnamefont{Mei{\ss}ner}}
  (\bibinfo{year}{in preparation}).

\bibitem[{\citenamefont{Gasser et~al.}(1991)\citenamefont{Gasser, Leutwyler,
  and Sainio}}]{Gasser:1990ce}
\bibinfo{author}{\bibfnamefont{J.}~\bibnamefont{Gasser}},
  \bibinfo{author}{\bibfnamefont{H.}~\bibnamefont{Leutwyler}},
  \bibnamefont{and} \bibinfo{author}{\bibfnamefont{M.}~\bibnamefont{Sainio}},
  \bibinfo{journal}{Phys.Lett.} \textbf{\bibinfo{volume}{B253}},
  \bibinfo{pages}{252} (\bibinfo{year}{1991}).

\bibitem[{\citenamefont{Pavan et~al.}(2002)\citenamefont{Pavan, Strakovsky,
  Workman, and Arndt}}]{Pavan:2001wz}
\bibinfo{author}{\bibfnamefont{M.}~\bibnamefont{Pavan}},
  \bibinfo{author}{\bibfnamefont{I.}~\bibnamefont{Strakovsky}},
  \bibinfo{author}{\bibfnamefont{R.}~\bibnamefont{Workman}}, \bibnamefont{and}
  \bibinfo{author}{\bibfnamefont{R.}~\bibnamefont{Arndt}},
  \bibinfo{journal}{PiN Newslett.} \textbf{\bibinfo{volume}{16}},
  \bibinfo{pages}{110} (\bibinfo{year}{2002}), \eprint{hep-ph/0111066}.

\bibitem[{\citenamefont{Alarc{\'o}n et~al.}(2012)\citenamefont{Alarc{\'o}n,
  Camalich, and Oller}}]{Alarcon:2011zs}
\bibinfo{author}{\bibfnamefont{J.}~\bibnamefont{Alarc{\'o}n}},
  \bibinfo{author}{\bibfnamefont{J.}~\bibnamefont{Camalich}}, \bibnamefont{and}
  \bibinfo{author}{\bibfnamefont{J.}~\bibnamefont{Oller}},
  \bibinfo{journal}{Phys.Rev.} \textbf{\bibinfo{volume}{D85}},
  \bibinfo{pages}{051503} (\bibinfo{year}{2012}), \eprint{1110.3797}.

\bibitem[{\citenamefont{Kryjevski}(2004)}]{Kryjevski}
\bibinfo{author}{\bibfnamefont{A.}~\bibnamefont{Kryjevski}},
  \bibinfo{journal}{Phys.Rev.} \textbf{\bibinfo{volume}{D70}},
  \bibinfo{pages}{094028} (\bibinfo{year}{2004}), \eprint{hep-ph/0312196}.

\bibitem[{\citenamefont{Vecchi}(2013)}]{Vecchi}
\bibinfo{author}{\bibfnamefont{L.}~\bibnamefont{Vecchi}}
  (\bibinfo{year}{2013}), \eprint{1312.5695}.

\bibitem[{\citenamefont{Lopez-Honorez et~al.}(2012)\citenamefont{Lopez-Honorez,
  Schwetz, and Zupan}}]{LopezHonorez:2012kv}
\bibinfo{author}{\bibfnamefont{L.}~\bibnamefont{Lopez-Honorez}},
  \bibinfo{author}{\bibfnamefont{T.}~\bibnamefont{Schwetz}}, \bibnamefont{and}
  \bibinfo{author}{\bibfnamefont{J.}~\bibnamefont{Zupan}},
  \bibinfo{journal}{Phys.Lett.} \textbf{\bibinfo{volume}{B716}},
  \bibinfo{pages}{179} (\bibinfo{year}{2012}), \eprint{1203.2064}.

\bibitem[{\citenamefont{Frandsen et~al.}(2012)}]{Frandsen:2012db}
\bibinfo{author}{\bibfnamefont{M.~T.} \bibnamefont{Frandsen}}
  \bibnamefont{et~al.}, \bibinfo{journal}{JCAP}
  \textbf{\bibinfo{volume}{1210}}, \bibinfo{pages}{033} (\bibinfo{year}{2012}),
  \eprint{1207.3971}.

\bibitem[{\citenamefont{Crivellin et~al.}(2014)\citenamefont{Crivellin,
  D'Eramo, and Procura}}]{CEP}
\bibinfo{author}{\bibfnamefont{A.}~\bibnamefont{Crivellin}},
  \bibinfo{author}{\bibfnamefont{F.}~\bibnamefont{D'Eramo}}, \bibnamefont{and}
  \bibinfo{author}{\bibfnamefont{M.}~\bibnamefont{Procura}}
  (\bibinfo{year}{2014}), \eprint{1402.1173}.

\bibitem[{\citenamefont{Gao et~al.}(2013)\citenamefont{Gao, Kang, and
  Li}}]{Gao:2011ka}
\bibinfo{author}{\bibfnamefont{X.}~\bibnamefont{Gao}},
  \bibinfo{author}{\bibfnamefont{Z.}~\bibnamefont{Kang}}, \bibnamefont{and}
  \bibinfo{author}{\bibfnamefont{T.}~\bibnamefont{Li}}, \bibinfo{journal}{JCAP}
  \textbf{\bibinfo{volume}{1301}}, \bibinfo{pages}{021} (\bibinfo{year}{2013}),
  \eprint{1107.3529}.

\bibitem[{\citenamefont{Okada and Seto}(2013)}]{Okada:2013cba}
\bibinfo{author}{\bibfnamefont{N.}~\bibnamefont{Okada}} \bibnamefont{and}
  \bibinfo{author}{\bibfnamefont{O.}~\bibnamefont{Seto}},
  \bibinfo{journal}{Phys.Rev.} \textbf{\bibinfo{volume}{D88}},
  \bibinfo{pages}{063506} (\bibinfo{year}{2013}), \eprint{1304.6791}.

\bibitem[{\citenamefont{B{\'e}langer
  et~al.}(2013{\natexlab{b}})\citenamefont{B{\'e}langer, Goudelis, Park, and
  Pukhov}}]{Belanger:2013tla}
\bibinfo{author}{\bibfnamefont{G.}~\bibnamefont{B{\'e}langer}},
  \bibinfo{author}{\bibfnamefont{A.}~\bibnamefont{Goudelis}},
  \bibinfo{author}{\bibfnamefont{J.-C.} \bibnamefont{Park}}, \bibnamefont{and}
  \bibinfo{author}{\bibfnamefont{A.}~\bibnamefont{Pukhov}}
  (\bibinfo{year}{2013}{\natexlab{b}}), \eprint{1311.0022}.

\end{thebibliography}
\end{document}